\documentclass[11pt,a4paper]{article}

\pdfoutput=1
\usepackage{jcappub}
\usepackage{graphicx}
\usepackage{multirow}
\usepackage{hyperref}
\usepackage{amsmath,amssymb}
\usepackage{graphicx}
\usepackage{bm}
\usepackage{latexsym}
\usepackage{epsfig}
\usepackage{psfrag}
\usepackage{ulem}
\usepackage{color}
\usepackage[dvipsnames]{xcolor}
\usepackage{subfigure}
\usepackage[utf8]{inputenc}  

\usepackage[utf8]{inputenc}
\usepackage[a4paper, total={7in, 9in}]{geometry}
\makeatletter
\gdef\@fpheader{}
\makeatother
\newcommand{\be}{\begin{equation}}
\newcommand{\ee}{\end{equation}}
\newcommand{\bea}{\begin{eqnarray}}
\newcommand{\eea}{\end{eqnarray}}

\allowdisplaybreaks

\usepackage{hyperref}  
\usepackage{natbib}    
\usepackage{etoolbox}  
\usepackage{doi}       

\makeatletter
\patchcmd{\@bibitem}{\ignorespaces}{\ignorespaces\printdoi}{}{}
\newcommand{\printdoi}{%
  \ifthenelse{\isundefined{\@doi}}{}%
    {\par\noindent\texttt{\href{https://doi.org/\@doi}{DOI: \@doi}}}%
}
\makeatother

\begin{document}

\title{DESI results: Hint towards coupled dark matter and dark energy}

\author[a]{Amlan Chakraborty,}
\author[b]{Prolay K. Chanda,}
\author[a]{Subinoy Das,}
\author[c]{and Koushik Dutta}
\affiliation[a]{Indian Institute of Astrophysics, Bengaluru, Karnataka 560034, India}
\affiliation[b]{Tata Institute of Fundamental Research, Homi Bhabha Road, Mumbai 400005, India}
\affiliation[c]{Department of Physical Sciences, Indian Institute of Science Education and Research Kolkata, Mohanpur-741 246, WB, India}

\emailAdd{amlan.chakraborty@iiap.res.in}
\emailAdd{prolay.chanda@tifr.res.in}
\emailAdd{subinoy@iiap.res.in}
\emailAdd{koushik@iiserkol.ac.in}







\abstract{We investigate a scenario where a dark energy quintessence field $\phi$ with positive kinetic energy is coupled with dark matter. With two different self-interaction potentials for the field and a particular choice of the coupling function, we show explicitly how the observable effective equation of state parameter $w_{\rm eff}$ for the dark energy field crosses the phantom barrier ($w_{\rm eff} = -1$) while keeping the equation of state of the quintessence field $w_\phi > -1$. With appropriate choices of parameters, $w_{\rm eff}$ crosses the phantom divide around redshift $z\sim 0.5$, transitioning from $w_{\rm eff} <-1$ in the past to $w_{\rm eff}>-1$ today. This explains DESI observations well. Our analysis reveals that the model remains consistent within the $2\sigma$ confidence intervals provided by DESI for several combinations of the scalar field parameters, highlighting its potential in explaining the dynamics of dark energy arising from a simple Yukawa-type long-range interaction in the dark sector. While the current findings offer a promising framework for interpreting DESI observations, future work, including a comprehensive Markov Chain Monte Carlo (MCMC) analysis, is necessary to constrain the parameter space further and strengthen the statistical significance of the results.   
}

\maketitle

\section{Introduction}

Over the past decade, precision observations have favored the standard flat $\Lambda$CDM cosmology, reinforcing the simple picture of a cosmological constant ($\Lambda$) driving accelerated expansion \cite{refId0}.
In recent years, as more and more high-precision data are coming up, a few anomalies like the Hubble anomaly, $\sigma_8$ anomaly \cite{Abdalla:2022yfr} may seem to pose strong challenges to $\Lambda$CDM paradigm. And very recently, baryon acoustic oscillation (BAO) measurements from the first-year results from the Dark Energy Spectroscopic Instrument (DESI) have injected new momentum into dark energy model building \cite{DESI:Cosmology, Reconstruction,PhysRevD.111.023532}.

First of all, along with several other combinations of data sets like Type Ia supernova (SN Ia) and Cosmic Microwave Background (CMB), if DESI data is fitted with a dark energy sector with a {\it constant} equation of state parameter $w$ (namely $w$CDM model), it is fully consistent with $w = -1$. This hints towards a cosmological constant. Relying on the poor theoretical understanding of the origin of the cosmological constant \cite{Martin:2012bt,Bamba:2012cp}, in particular, its required magnitude, the data is fitted with a dark energy sector, which is dynamical \cite{DESI:Cosmology}. Considering that the origin of the dark sector is still unknown, it was parameterized by two constant parameters (known as CPL parameterization \cite{Chevallier:2000qy, Linder:2002et}) $w_0$ and $w_a$ with the equation of state of dark energy is given by $w (a) = w_0 + w_a (1 -a)$, where $a$ is the scale factor representing the expansion of the Universe. Depending on various SN Ia data combinations with CMB, DESI data found that $w_0 > -1$ and $w_a < 0$ with a significance varying from $2.5 \sigma$ to $3.9 \sigma$ \footnote{Interpreting DESI results in terms of a specific parameterization has been criticized at \cite{Shlivko:2024llw, Cortes:2024lgw, Gialamas:2024lyw,Huang:2025som,RoyChoudhury:2024wri,Colgain:2024mtg,Chan-GyungPark:2025cri,Chan-GyungPark:2024brx, Banerjee:2020xcn, Lee:2022cyh}. }. This interpretation of data means that
$\Lambda$CDM is disfavored
at $\sim 2.5–3.9 \sigma$. Moreover, the above constraint on $w_0$ and $w_a$ would mean that the dark energy sector had an equation of state $w < -1$ for $z > 1$ that crossed $w = -1$ phantom barrier at some recent past with its present value $w_0 > -1$ \cite{Linder:2024rdj}. To bolster the above arguments, the DESI collaboration reconstructed the equation of state of the dark energy sector by writing its equation of state $w(z)$ in terms of orthogonal polynomials \cite{Reconstruction}. It was reported that the data favors evolving dark energy with cosmological constant lying outside $2$-$\sigma$ confidence contour for some redshift ranges. Moreover, the findings agree with the interpretation in terms of $w_0$ and $w_a$. A similar conclusion, albeit depending on the dataset used, was drawn in \cite{Mukherjee:2024ryz, Jiang:2024xnu, Ye:2024ywg,Wolf:2025jlc,Wolf:2024stt,Ferrari:2025egk}. Also, non-DESI data has been used to confirm and strengthen the DESI 2024 \cite{Chan-GyungPark:2024mlx}. Moreover, DESI has been tested against several other parameterizations than CPL, and all lead towards dynamical dark energy \cite{Giare:2024gpk}. 
In short, DESI collaboration results hint towards tantalizing evidence for evolving dark energy that shows phantom crossing \footnote{Within three days of the manuscript's publication on ArXiv, the results from the DESI DR2 dataset were released \cite{DESI:2025zgx}, providing significantly stronger evidence for an evolving dark energy. The model-independent reconstruction of the dark energy equation of state with the DR2 dataset yields a $2$-$\sigma$ bound \cite{Lodha:2025qbg} that remains consistent with their findings from DESI DR1, indicating compatibility of our model (to be discussed in the next section) with the newly released dataset.}. 


Various dark energy models have been explored in light of possible departures from \( w=-1 \)\cite{Odintsov:2024woi}. A well-studied class is quintessence \cite{Copeland:2006wr, Ratra:1987rm, PhysRevLett.82.896}, a slowly rolling scalar field producing time-varying dark energy pressure. In quintessence models, the dynamics fall into two broad categories: freezing, where the field slows and approaches \( w \to -1 \), and thawing, where it begins frozen at \( w \approx -1 \) and only recently started evolving \cite{PhysRevLett.95.141301, PhysRevD.73.063010,PhysRevD.91.063006,Berbig:2024aee}. While many potentials, exponential, power-law, etc., have been studied, they all yield \( -1 \leq w < -1/3 \) \cite{Tsujikawa_2013,Cahn_2008,Bhattacharya:2024hep}, ensuring accelerated expansion but preventing a crossing into the phantom regime, (\( w < -1 \)), thus not suitable to explain the DESI anomaly \cite{Wolf:2024eph, Ramadan:2024kmn,Bhattacharya:2024kxp}.
Quintessential interpretations of the evolving dark energy in light of DESI observations have been looked at, but the dark energy eventually becomes phantom-like in the past \cite{Tada:2024znt}.

Meanwhile, phantom dark energy models \cite{CALDWELL200223} extend into \( w < -1 \) but at the cost of violating the null energy condition, requiring exotic physics such as a negative kinetic term \cite{doi:10.1142/S0217732317300257,PhysRevLett.91.071301}. The models where dark energy originates from a phase transition like vacuum metamorphosis \cite{PhysRevD.62.083503,PhysRevD.73.023513,PhysRevD.97.043528}, or ``quintom" scenarios\cite{GUO2005177,Yang:2025kgc}, struggles to explain the DESI anomaly either from an observational point of view or from model building aspects, pointing to the need for new or more complex dynamics \cite{PhysRevD.111.023532,SHLIVKO2024138826,Fikri:2024klc,PhysRevD.110.083528,HERNANDEZALMADA2024101668, CAI20101,Yin:2021uus,Yin:2024hba,Heckman:2024apk}. However, it has been shown that using a modified gravity model (e.g.a general Horndeski scalar-tensor theory) can provide a viable explanation for the DESI observations \cite{PhysRevD.110.123524, Ye:2024ywg,Ferrari:2025egk,Chudaykin:2025gdn}.

Here in this paper, we focus on interactive dark energy models from the perspective of DESI results.
Interacting dark energy models have also been an important extension of the dark energy paradigm by allowing energy exchange between dark energy and dark matter, a possibility motivated by theoretical arguments that a scalar field naturally couples to other sectors unless protected by symmetry \cite{PhysRevD.62.043511, Amendola:2002bs, Amendola:2003eq, Amendola:2000ub}. Such models were initially proposed to address the coincidence problem and exhibit distinctive observational signatures, including modified expansion and structure growth rates. Recent analyses indicate that a non-gravitational interaction in the dark sector is still consistent with current observations \cite{Chakraborty:2024xas,Gomez-Valent:2020mqn,Gomez-Valent:2022bku,Goh:2022gxo}and might even help relieve tensions in the standard $\Lambda$CDM model \cite{Wang_2024,Cai:2021wgv,Benisty:2022lox,Benisty:2021cmq,Benisty:2023dkn,Sabogal:2024yha, Shah:2024rme, Shah:2025ayl}. Notably, a joint analysis of DESI Year 1 BAO and Planck 2018 CMB data shows a preference for dark sector interaction at greater than $95\%$ confidence, with a coupling parameter shifting $H_0$ toward the locally measured value \cite{PhysRevLett.133.251003}, offering an intriguing hint of interacting dark sector \cite{Lewis:2024cqj,universe11010010,Li:2025owk,Li_2024, Li:2024qso,Sabogal:2025mkp} frameworks alongside more conventional models. Implications of DESI results for the Hubble tension have also been explored in \cite{Wang:2024dka}.

In this work, we propose an interacting dark sector model, especially a chameleon model \cite{PhysRevD.69.044026,Paliathanasis:2024sle,Paliathanasis:2020wjl}, as an economical solution to DESI anomaly. By allowing dark energy to exchange energy with dark matter, our framework naturally mimics phantom-like behavior without invoking a fundamental phantom field \cite{Das:2005yj}. 
We introduce an interaction term modifying the effective equation of state while maintaining theoretical stability. Unlike standard chameleon models \cite{Das:2005yj,PhysRevD.69.044026,Paliathanasis:2023dfz,Paliathanasis:2023ttu}, where the scalar field settles at a potential minimum with vanishing kinetic energy, our approach allows for continued evolution, ensuring non-zero kinetic energy at present. This setup accommodates an evolving history, permitting early phantom behavior followed by a transition to quintessential behavior today, which is consistent with DESI findings. The two most important points in our paper are the following:
\begin{itemize}
\item Due to the interaction between dark matter and dark energy, the inferred equation of state by an observer is different from the equation of state of the scalar field $w_{\phi}$.
\item Along with the interactions, the finite kinetic energy of the scalar field at the present epoch allows the effective (observed) equation of state to cross the phantom barrier.
\end{itemize}

The plan of the paper is as follows. In Section \ref{sec: Model}, we discuss the Chameleon model framework and the mechanism invoked to explain the DESI observation. In Section \ref{sec: Model_Imp}, we present the numerical implementation of the said model and discuss the obtained results in Section \ref{Sec:result}. Finally, we conclude in Section \ref{sec:conclusions}.





\section{Model Set Up}
\label{sec: Model}
In the section, we revisit the model proposed in \cite{Das:2005yj} where the quintessence dark energy model of a scalar field $\phi$ with canonical {\it positive} kinetic energy term interacts with dark matter in Einstein gravity. We will argue that the model can easily {\it mimic} effective (observable) dark energy equation of state parameter $w(z)$ that crosses the phantom divide in the recent past and reaches its present value $w_0 > - 1$. Therefore, the setup is ideally suited to explain the observations made by DESI. We will show explicit examples of dark energy sectors and how they fit the data. 

This study considers a quintessence scalar field $\phi$, representing dark energy, which has a Yukawa-like interaction with the dark matter (fermion) field $\psi$ \cite{Das:2005yj} \footnote{Here, the nature of the dark matter is not so important. The setup could well be generalized to bosonic dark matter with different interaction terms.}, 
\begin{equation} \label{intercations}
{\mathcal L}_{\rm int} = - f(\phi/M_{\rm Pl})\bar{\psi}\psi~.
\end{equation}
We note that the interaction term makes the dark matter mass quintessence field dependent\footnote{ Note that the dark energy field is coupled only with the dark matter. Although moduli fields can in principle couple universally to all matter sectors, we assume that the visible sector exhibits a Damour–Polyakov least-coupling mechanism \cite{Damour:1994zq}, which dynamically suppresses baryonic interactions and keeps them within equivalence-principle bounds. The dark sector, however, can retain a non-vanishing coupling, as expected in sequestered or hidden-sector realizations.}. In effect, the non-relativistic dark matter energy density does not redshift as $a^{-3}$ as for the case of ordinary matter, but evolves as follows.
\begin{equation}
\rho_{\rm DM} \sim f(\phi)/a^3~,
\end{equation}
where, $a$ is the scale factor, with the present value of the scale factor normalized as $a_0 = 1$, and the field is denoted as $\phi_0$. We begin with a general form of $f(\phi)$, but observational requirements will fix the properties of the function, as we will see later.

In this case, the Hubble parameter is determined by the following expression. 
\begin{equation} \label{Hubble_1}
3H^2M_{\rm Pl}^2 = \rho_{\phi} + \frac{\rho_{\rm DM}^{(0)}}{a^3}\frac{f(\phi)}{f_0}+\frac{\rho_{B}^{(0)}}{a^3},
\end{equation}
where, $\rho_{\rm DM}^{(0)}$ and $\rho_{B}^{(0)}$ are the present-day dark matter density and baryon density, respectively, and $f_0 = f(\phi_0)$, and $M_{\rm Pl}$ is the reduced Planck mass. The dark energy density is given as the sum of the kinetic and potential energy density of the quintessence scalar field. 
\begin{equation}
\rho_{\phi} = \frac{\dot \phi^2}{2} + V(\phi)~.
\end{equation}

\begin{figure}
\centering
\includegraphics[scale = 0.6]{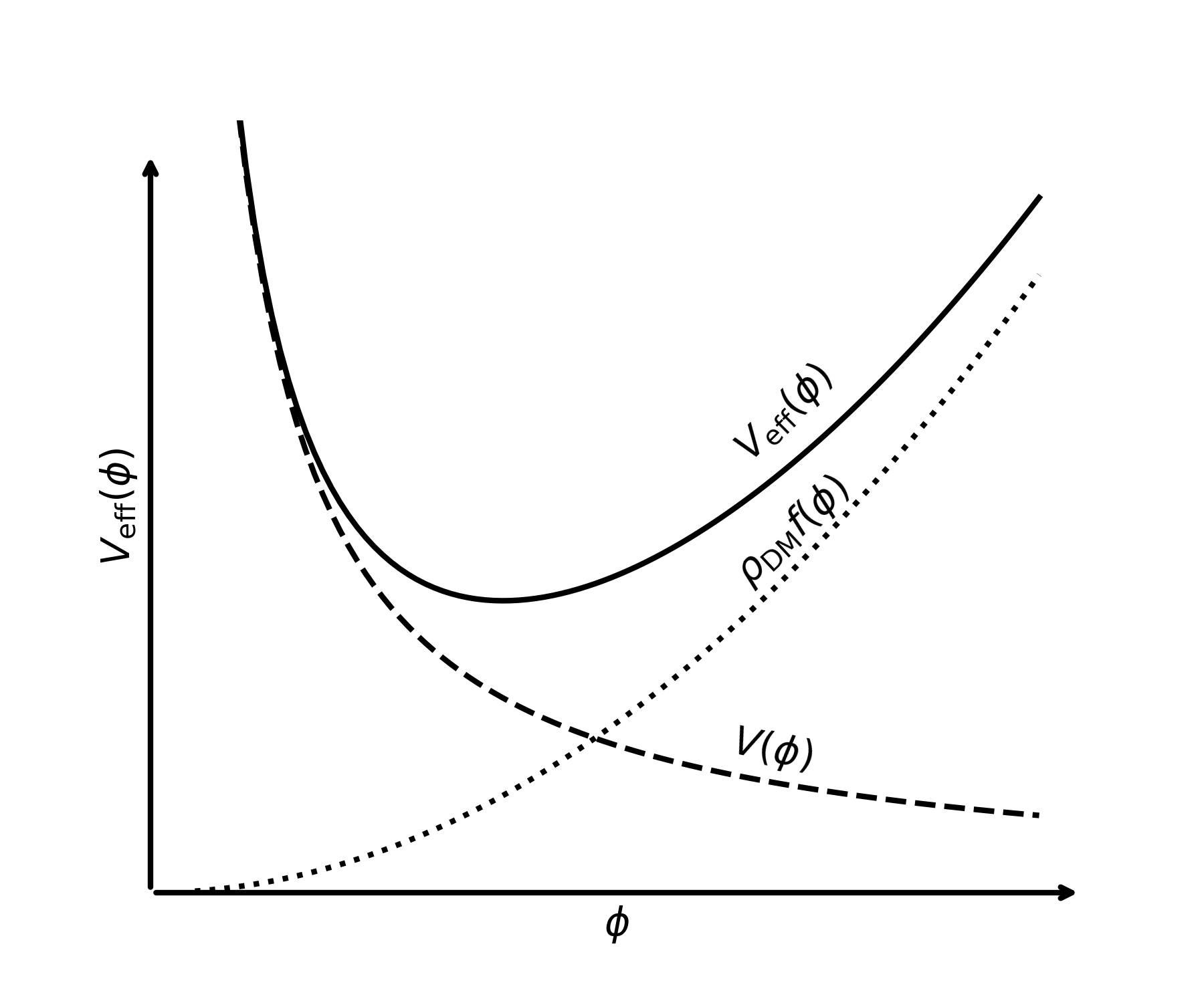}
\caption{Schematic plot of \(V_{\text{eff}}(\phi) = V(\phi) + \rho_{\rm DM}f(\phi)\).}
\label{fig:Veff}
\end{figure}

By varying the action related to $\phi$ we obtain the following modified Klein-Gordon (KG) equation of motion,
\begin{equation}
  \ddot{\phi} + 3H\dot{\phi} = -V_{,\phi}(\phi) - \frac{\rho_{\rm DM}^{(0)}}{a^3} \frac{f_{,\phi}(\phi)}{f(\phi_0)}~,
 \label{eq_KG_phi}
\end{equation}
where interaction between dark matter and the quintessence field invokes the last term and `$,\phi$' in the subscript to any quantity corresponds to the derivative of the quantity w.r.t. $\phi$. The above equation tells us that the dynamics of the field are determined by an effective potential, $V_{\rm eff} = V(\phi)+\rho_{\rm DM}^{(0)}f(\phi)/a^3 f_0$. This is schematically shown in Fig. \ref{fig:Veff} where $V(\phi)$ is a falling function of $\phi$ and $f(\phi)$ is a growing function of $\phi$. Two competing behaviors of the functions create a minimum of the potential. It is important to note that the minimum of the potential moves as the strength of the growing function determined by dark matter density dilutes with the expansion of the Universe. Depending on the choices of these two functions and initial conditions, the quintessence field may either sit at the minimum of the effective potential or show some additional dynamics around the minimum. In the first case, the effective mass of the field being heavy, the field tracks the minimum adiabatically. We will see later that to mimic DESI results, the field must not sit at its effective minimum and, therefore, violate the adiabatic condition. 

At this moment, we note that the total Hubble parameter
and its integrals determine the cosmological observables. Therefore, any inference on dark energy properties from cosmological observables is only indirect. In our case, an observer would be unaware of the interactions between dark matter and dark energy. Any individual observer will infer the properties of dark energy using the following expressions for the same Hubble equation of \eqref{Hubble_1}
\begin{equation}
3H^2 M_{\rm Pl}^2 = \rho_{\rm DE}^{\rm eff} + \frac{\rho_{\rm DM}^{(0)}}{a^3}+ \frac{\rho_{\rm B}^{(0)}}{a^3}~.
\end{equation}
Here, $\rho_{\rm DE}^{\rm eff}$ satisfies the energy conservation equation
\begin{equation}
\frac{d}{dt} (\rho_{\rm DE}^{\rm eff}) + 3H (1 + w_{\rm eff}) \rho_{\rm DE}^{\rm eff} = 0~,
\end{equation}
where, $w_{\rm eff}$ is the effective equation of the state of the dark energy as perceived or measured by an observer. Take note that the above two equations are written in a way such that there seems to be no interaction between the dark matter (represented by $\rho_{\rm DM}$) and the dark energy (represented by $\rho_{\rm DE}^{\rm eff}$). 

It has been shown \cite{Das:2005yj} that the effective equation of state of the dark energy can be written as
\begin{equation}\label{eq:w_eff}
  w_{\rm eff} = \frac{w_\phi}{1 - x}~,
\end{equation}
where
\begin{equation}\label{eq:x}
 x = - \frac{\rho_{\rm DM}^0}{a^3 \rho_\phi} \left(\frac{f(\phi)}{f_0} - 1\right)~,
\end{equation}
and 
\begin{equation}\label{eq:w_phi}
w_{\phi} = \frac{\left(\dot{\phi}^2/2 - V(\phi)\right)}{\left(\dot{\phi}^2/2 + V(\phi)\right)}.
\end{equation}
is the usual equation of state parameter for a dark energy field with a canonical kinetic energy term. We consider dynamics of $\phi$ such that $f(\phi)$ is increasing with time, and it ensures that $x>0$. In the case of nearly flat potentials, $w_\phi\approx -1$, and, therefore, in the past, observationally inferred $w_{\rm eff}$ can have values smaller than $-1$. Note that this requires a phantom field that needs the kinetic energy to be negative. Moreover, the setup has broad structural flexibility to cross $w_{\rm eff} < -1$ in the recent past by appropriately adjusting $w_{\phi}$ and $x$.

Let us first note that if the dark energy sector is modeled by a cosmological constant with $w_\phi = -1$ for all time and the coupling to the dark matter sector is also non-dynamical, it means $x = 0$. That leads to $w_{\rm eff} = -1$ for all the time. This is not suitable for explaining the DESI results. On the other hand, if a thawing quintessence field models the dark energy sector, the field starts to roll as the Universe expands, and it was stuck at its potential at earlier times {\cite{PhysRevLett.95.141301, Tsujikawa_2013, PhysRevD.73.063010,PhysRevD.91.063006}. As representative examples, if $w_\phi$ changes from $-1$ to $-0.8$ with an intermediate value of $-0.5$ while $x$ changes from $0.9$ to $0$ while $x = 0.5$ at the intermediate value, it can be easily checked from Eq.~\eqref{eq:w_eff} that the $w_{\rm eff}$ shows phantom barrier crossing in some recent past with $w_{\rm eff} > -1$ today. At the same time, the behavior of phantom crossing can also be reproduced when $w_\phi$ changes from $-1$ to $-0.9$ to $-0.8$ while $x$ changes from $0.9$ to $0.1$ to $0$. Note that in the first case, the field crosses its minimum and thus reduces the value of $w_\phi$ as the field loses its kinetic energy. In the second case, the field does not cross the minimum. It is instructive to note that the above example points are intended to illustrate, in principle, the flexibilities of the setup. Later, we will show explicit numerical examples that reproduce DESI results of $w_{\rm eff}$.

Our investigation is especially motivated by theoretical insights derived from String theory compactifications, which propose a relationship between the cosmological scalar field and the geometry of compactified dimensions \cite{Damour:1994zq}. In pursuit of this, we have introduced a coupling function, denoted $f(\phi)$, to mathematically represent these interactions, 
\begin{align} \label{coupling}
f(\phi) = \exp\left(\frac{\beta\phi}{\sqrt{8\pi }M_{\rm Pl}}\right),
\end{align}
where $\beta$ represents the strength of the coupling. This approach is similar to that utilized by \cite{Das:2005yj}, but has been specifically adapted to align with the intricacies of our requirements to suit the DESI data. Regarding the self-interaction potential of the scalar field, we have chosen a polynomial form :
\begin{equation}\label{eq_vphi_1}
   V(\phi) = M^4 \left(\frac{M_{\rm Pl}}{\phi}\right)^{\alpha}~,
\end{equation}
and an exponential form expressed as
\begin{equation}
   V(\phi)=M^4 \exp\left({-\frac{\alpha\phi}{\sqrt{8\pi}M_{\rm Pl}}}\right)~.
\label{eq_vphi_2}
\end{equation}
Here, the parameter $M$ has a dimension of energy, and the $\alpha$ parameter governs the steepness of the potential. Note that a combination of Eq.~\eqref{coupling} and one of the expressions of Eq.~\eqref{eq_vphi_1} or Eq.~\eqref{eq_vphi_2} produces the desired result of Fig.~\ref{fig:Veff}.

The scalar field evolves under the influence of an effective potential towards a dynamically evolving minimum. Prior studies in \cite{Das:2005yj} have established that the effective equation of state parameter $w_{\rm eff}$ tends to $-1.0$ in the present cosmological epoch. This trend is attributed to the specific setup, in which the scalar field reaches equilibrium at the minimum of the potential with $w_\phi=-1$ at the present epoch with $x=0$. However, this alignment with $w_{\mathrm{eff}}=-1$ does not fully accord with recent observations from DESI \cite{DESI:Cosmology}. To remedy the situation, our work introduces a nuanced revision to the above framework. Our hypothesis suggests that by allowing the scalar field not to stop at the potential minimum at the present epoch, we endow our model with the versatility necessary to better correlate with the findings from the DESI survey. This adjustment allows the equation of the state parameter, $w_{\rm eff}$, to exceed the limit $-1$, thus providing a credible interpretation of recent observational evidence. Through this refined methodology, we endeavor to bridge the gap between the evolving theories of dark energy and the most current empirical data.

\section{Model Implementation} \label{sec: Model_Imp}

In exploring scalar field cosmological models, the choice of exponential and polynomial potential presents significant challenges, especially in determining the final scalar field value analytically, $\phi_0$. Since our model does not require $\phi_0$ to coincide with the effective potential minima to match DESI observations, we adopt arbitrary values for $\phi_0$, along with the parameters $\alpha$ and $\beta$. This model combines thawing and interaction scenarios, where the field stays frozen at its initial position from an early time and begins rolling at late times ($a=0.2$), leaving the early universe cosmology largely unchanged.

In our study, we have numerically solved the Klein-Gordon (KG) equation \ref{eq_KG_phi} using a Python code, appropriately calculating the Hubble parameter at each epoch through the Friedmann equation \ref{Hubble_1} to understand the late-time dynamic behavior of the scalar field. We appropriately fix the parameters $M$ and $\phi_0$ as well as $\alpha$ and $\beta$. Instead of employing automated optimization methods like the shooting algorithm, we manually set the initial conditions for the field’s value and velocity for a fixed $M$ and $\phi_0$. We fix the initial velocity of the field to be zero and vary only the initial position of the field such that the present-day position of the field matches with $\phi_0$, within an accuracy of $0.01$ and at the same time, the corresponding kinetic and potential energy of the field adds up to the observed present-day dark energy density, $\rho_{\rm DE}$, within a level of precision of $0.01$. Following the obtained dynamics of the scalar field, i.e., its position and velocity for each epoch, as depicted in Figure \ref{fig:enter-phi_phi_dot_dyna}, we proceed to calculate the equation of state parameter for the field, $w_\phi$, using equation \ref{eq:w_phi}. We also determined the effective equation of state, $w_{\mathrm{eff}}$, by applying equations \ref{eq:x} and \ref{eq:w_eff}, with the results presented in Figure \ref{fig:fitted_recons_pol} \& \ref{fig:fitted_recons_exp}. To assess the validity of our model, we compared the $2-\sigma$ bound on the reconstructed dark energy equation of state $w(z)$, inferred from DESI observations \cite{Reconstruction}, against $w_{\mathrm{eff}}(z)$. We repeat the whole process by varying the parameters $\alpha, \beta, M, \phi_0$, until the obtained $w_{\mathrm{eff}}(z)$ properly fits within the $2-\sigma$ bound provided by DESI.

Our methodology follows the formalism established by \cite{Das:2005yj} where the kinetic energy of the scalar field is set to zero upon reaching the potential minima (i.e., $\phi_0$ in their case) in the present-day universe, making its potential energy the sole contributor to the dark energy density. However, our model introduces more flexible dynamics of the scalar field. Choosing the model parameters suitably and letting the field be at an arbitrary position in the present epoch permits the scalar field to efficiently cross the minima of the effective potential, reducing the self-interacting potential energy relative to the observed dark energy density. Consequently, this difference in energy density can be attributed to the scalar field's kinetic energy.

\begin{figure}
    \centering
    \includegraphics[width=0.53\linewidth]{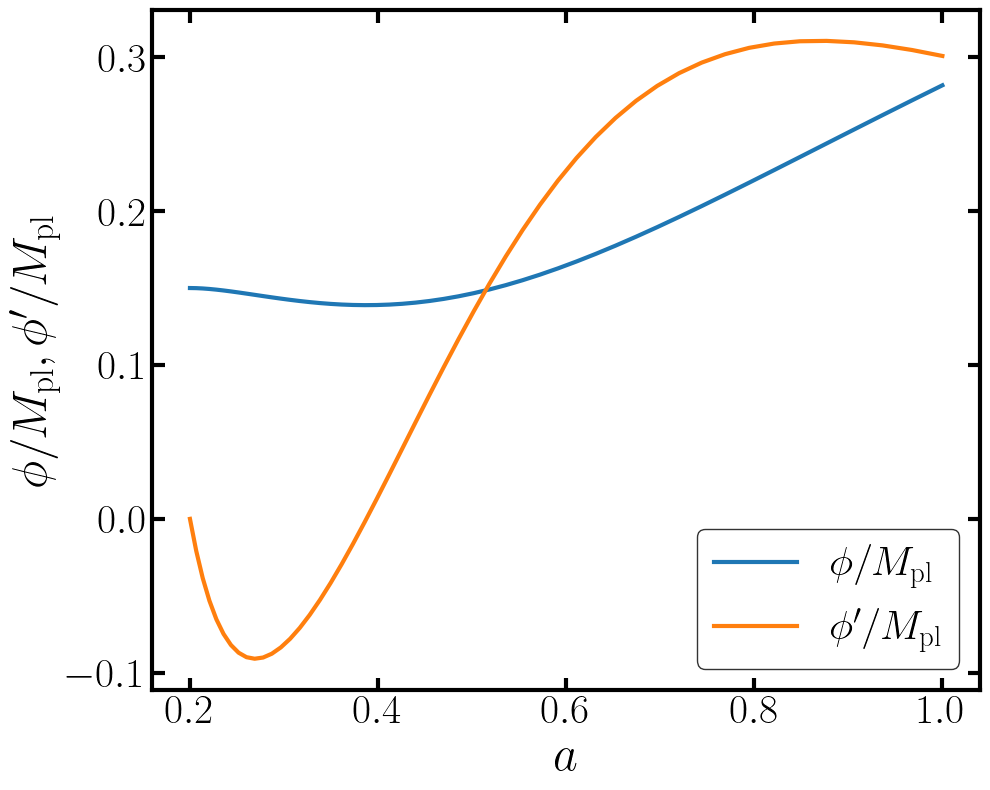}~\includegraphics[width=0.53\linewidth]{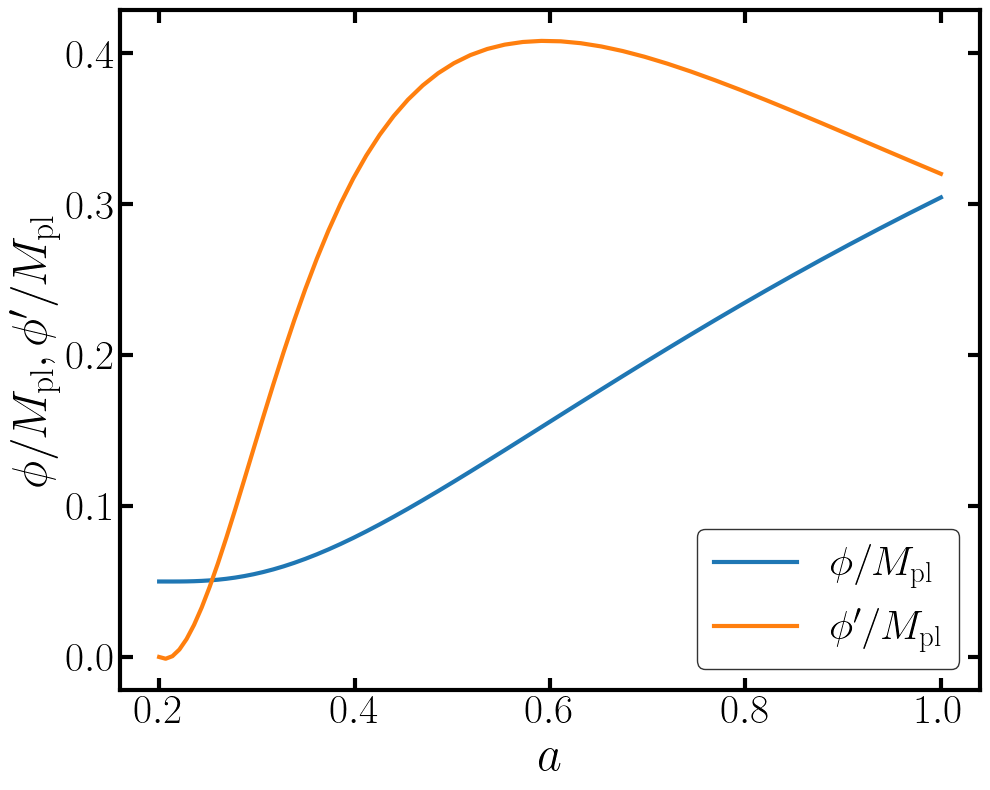}
    \caption{A typical dynamics of $\phi$ and $\phi^\prime=\frac{d\phi}{da}$ with the scale factor for the polynomial self-interaction potential is shown here with different initial positions of the field. The left (right) figure depicts the dynamics for a field that starts on the right (left) side of the effective minimum for $\alpha=0.14$ and $\beta=0.45$, with initial field value $\phi_{\rm initial}=0.15 M_{\rm pl} (0.05 M_{\rm pl})$ and final value $\phi_0=0.28 M_{\rm pl} (0.30 M_{\rm pl})$. Here $M_{\rm pl}$ is the reduced Planck mass.}
    \label{fig:enter-phi_phi_dot_dyna}
\end{figure}

Next, the late-time dynamics are governed by an effective potential arising from the self-interaction of the field and its coupling to dark matter. This creates an evolving effective minima that the field is attracted to over time. For $\alpha=0.14$, $\beta=0.45$, $M=2.14\times10^{-3}~ \text{eV}$, and $\phi_0=0.28 M_{\rm pl}$, starting from a stationary point on the right side of the potential, with $\phi_{\rm initial}=0.15 M_{\rm pl}$, as shown in the left panel of Figure \ref{fig:enter-phi_phi_dot_dyna}, the field rolls down toward the minimum, leading to a decrease in both its position and velocity. As the minima evolves to a higher value, the field starts to roll up the left side of the potential upon crossing it, eventually losing all of its kinetic energy. Subsequently, it starts to move back towards the minima, such that the field value and its velocity continue to increase until it crosses the minima again, showcasing an oscillatory pattern in the scalar field dynamics. As another example, for the right panel of Figure \ref{fig:enter-phi_phi_dot_dyna} with the same choice of $\alpha, \beta$ and $M$, the field starts rolling from the left side of the potential, with $\phi_{\rm initial}=0.05 M_{\rm pl}$, moves towards the evolving minima and eventually crosses it with an ever-increasing field value to reach the chosen $\phi_0=0.30 M_{\rm pl}$ in this case. The velocity, however, increases until the field crosses the minima and then decreases. 

\section{Results}
\label{Sec:result}

In this section, we show the results for the interacting quintessence dark matter scenario described by Eq.~\eqref{intercations}, Eq.~\eqref{coupling}, and Eq.~\eqref{eq_vphi_1} or Eq.~\eqref{eq_vphi_2}. In particular, we solve the dynamics of the quintessence field following the procedure outlined in the previous section and plot $w_{\rm eff}$ and $w_\phi$ as a function of redshift $z$ and confront them with the DESI results. We show illustrative examples of parameters for which predictions fall within the $2$-$\sigma$ bounds. 

The DESI collaboration has reconstructed the dark energy equation of state by expanding it in Chebyshev polynomials up to fourth order and constrained its coefficients via Monte Carlo Markov Chain (MCMC) analysis \cite{Reconstruction}. It provided strong evidence for the late-time dynamical behavior of dark energy. The $2$-$\sigma$ confidence interval they obtained from various datasets has been crucial for our study. Rather than conducting a separate MCMC analysis, we focus on verifying whether our prediction from the model is consistent with the observation from DESI and other key datasets (e.g., Union3, Planck). At the moment, it is important to remind ourselves that $w_{\rm eff}$ (not $w_{\phi}$) must be compared with the data.

We first focus on the model described by the potential given by Eq.~\eqref{eq_vphi_1}. This model combines thawing quintessence and interaction scenarios, where the field stays frozen at its initial position from an early time and begins rolling at late times, leaving the early universe cosmology largely unchanged. The results are illustrated in Fig.~\ref{fig:fitted_recons_pol} for different values of the model parameters. The shaded areas indicate the $2$-$\sigma$ confidence region as per various datasets, as provided in \cite{Reconstruction}. In all cases, the continuous violet line corresponds to $w_{\rm eff}$, where the blue dashed line delineates $w_\phi$. Furthermore, we include a reference line at $w=-1$ to highlight the significance of phantom crossing, which is crucial for interpreting the DESI observation \cite{DESI:Cosmology, Reconstruction}. Note that $w_{\phi}$ never crosses the phantom barrier, but the observable $w_{\rm eff}$ does.

\begin{figure}
  \centering
    \includegraphics[width=0.5\linewidth]{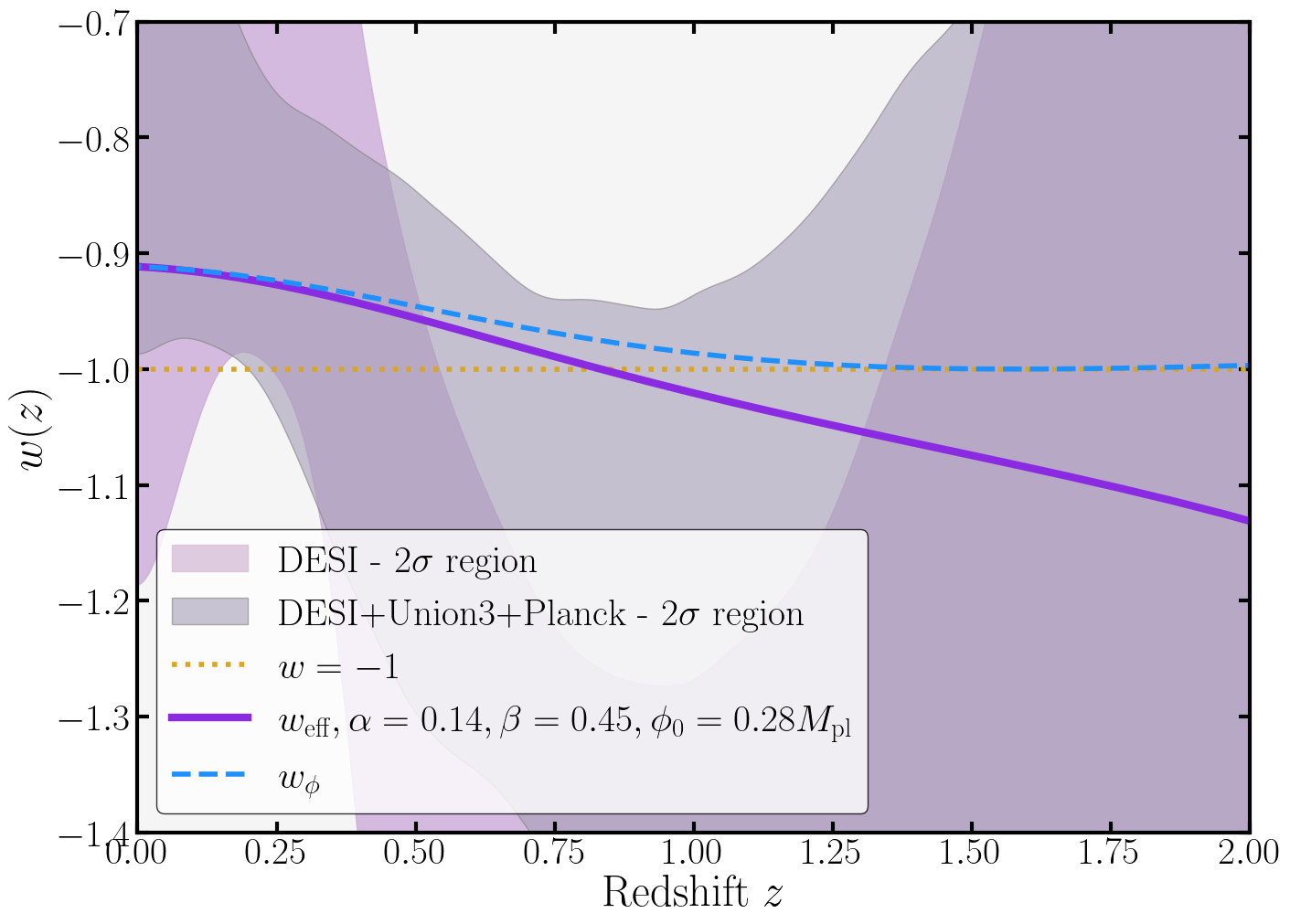}~\includegraphics[width=0.5\linewidth]{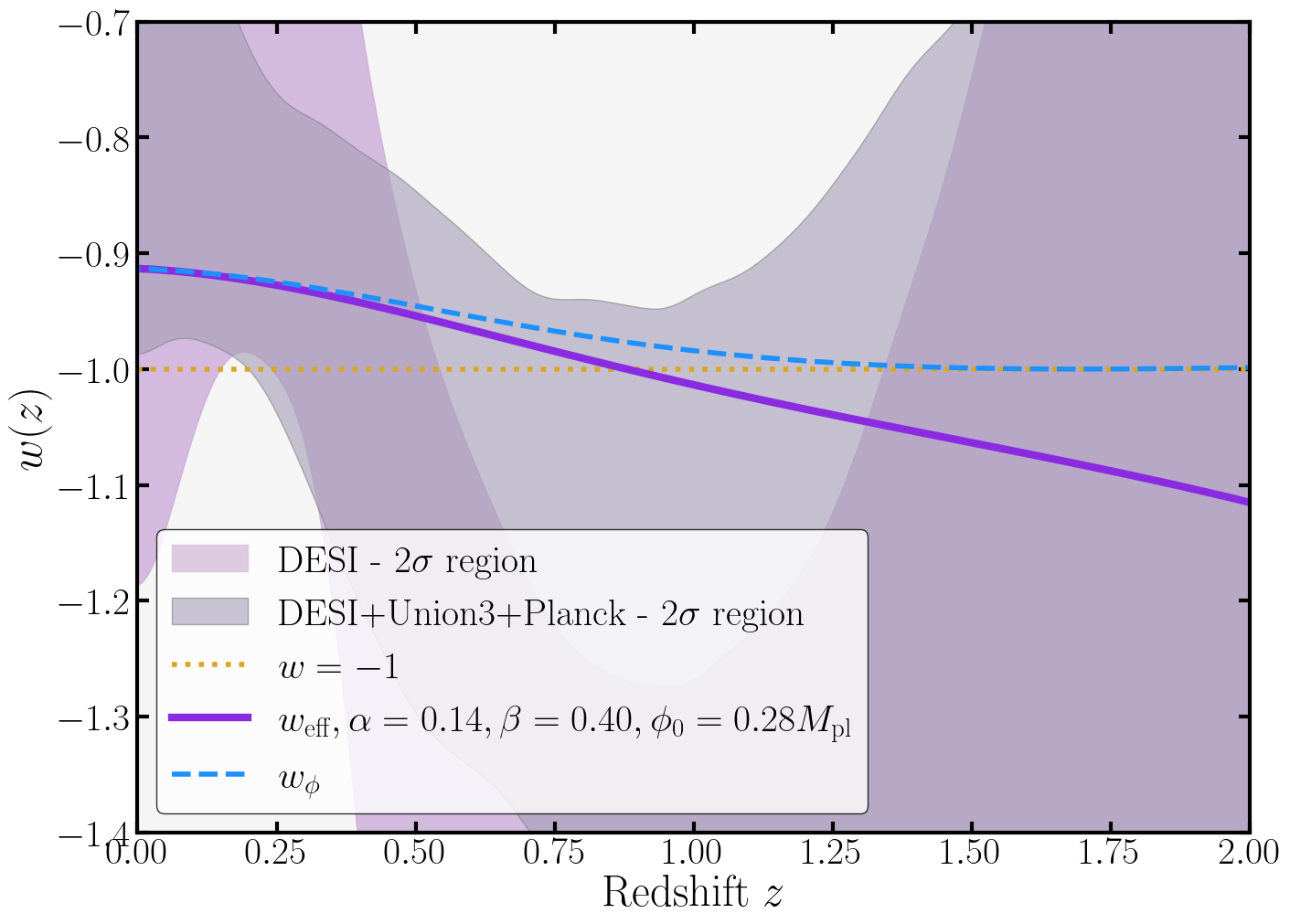}

\includegraphics[width=0.5\linewidth]{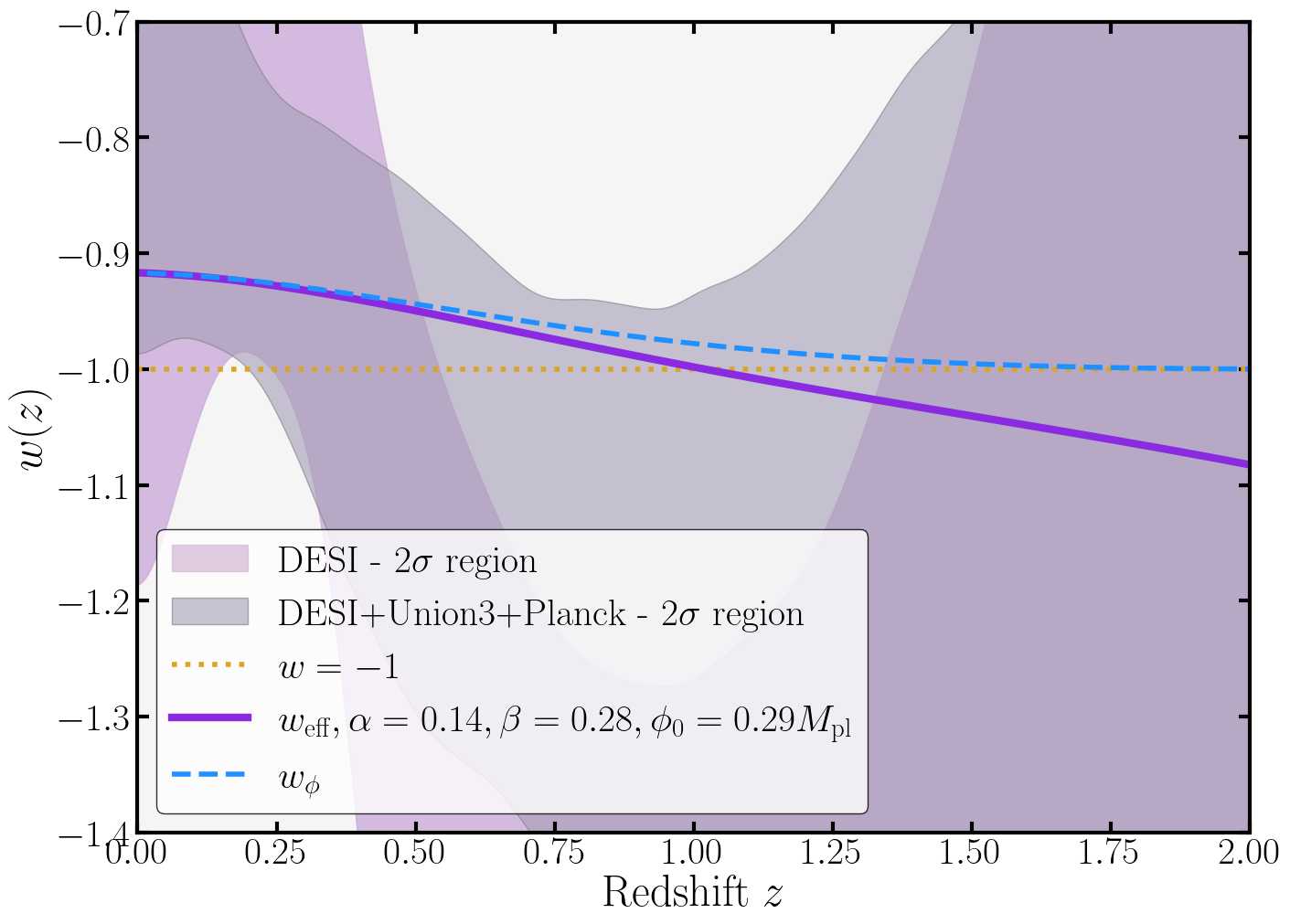}~\includegraphics[width=0.5\linewidth]{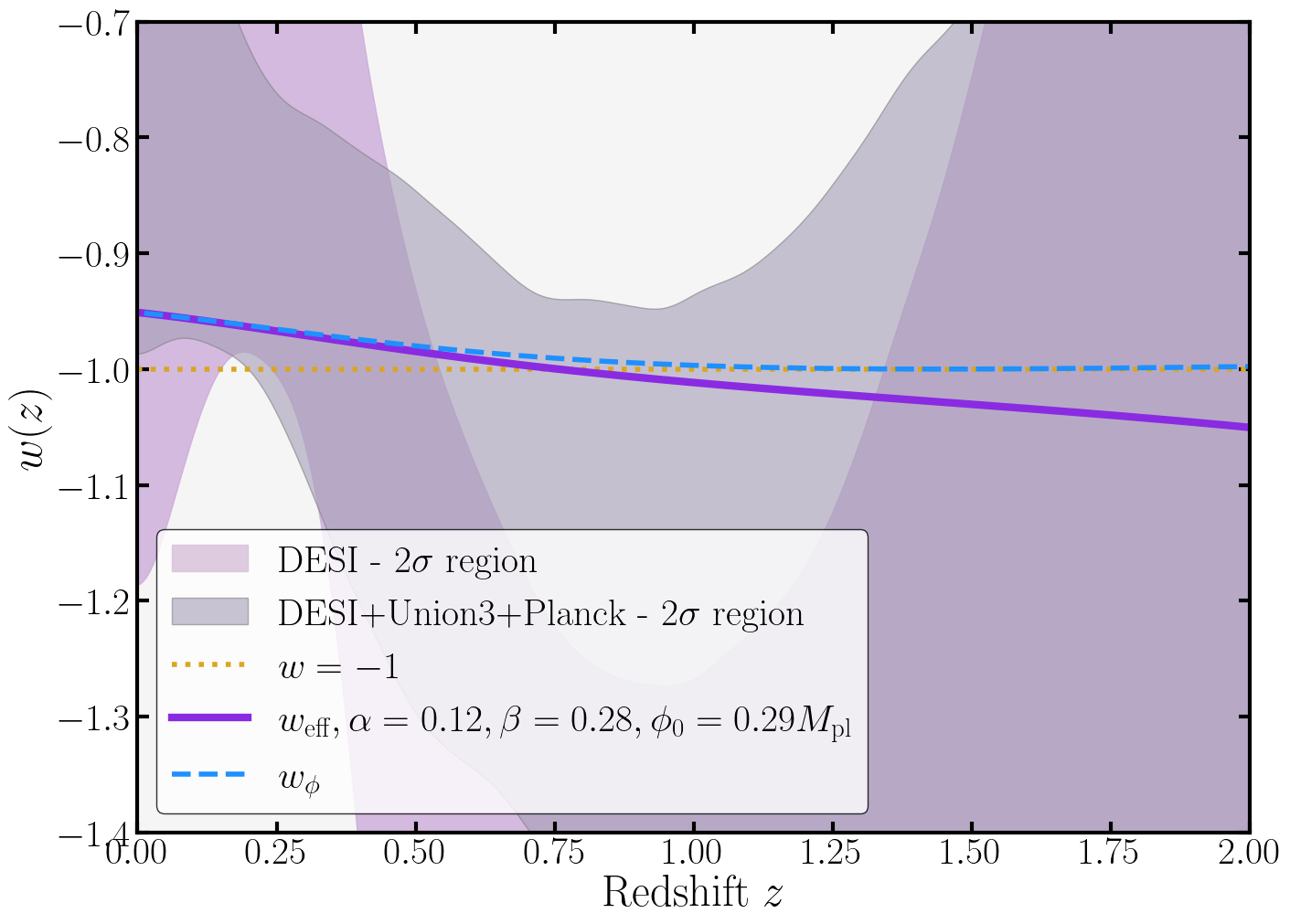}
 \caption{Fitting of $w(z)$ for polynomial self-interaction potential within the $2$-$\sigma$ contours of DESI (light purple) and DESI+Union3+Planck (light grey). The initial value of phi, $\phi_{\rm in}$ is considered as $0.15 M_{\rm pl}$(top-left), $0.15 M_{\rm pl}$ (top-right), $0.15 M_{\rm pl}$ (bottom-left) and $0.21 M_{\rm pl}$ (bottom-right).}
 \label{fig:fitted_recons_pol}
\end{figure}

\begin{figure}
 \centering
 \includegraphics[width=0.5\linewidth]{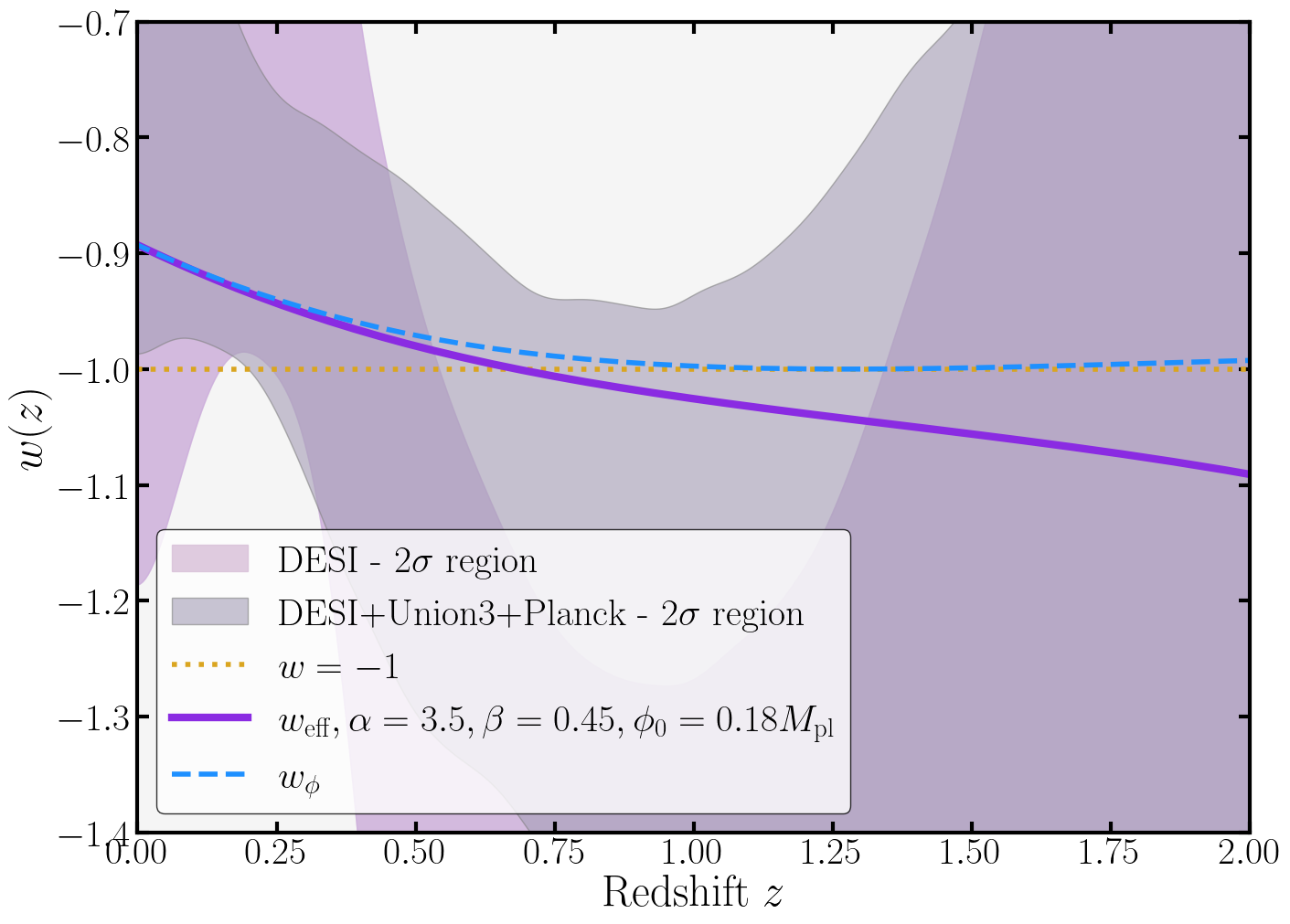}~\includegraphics[width=0.5\linewidth]{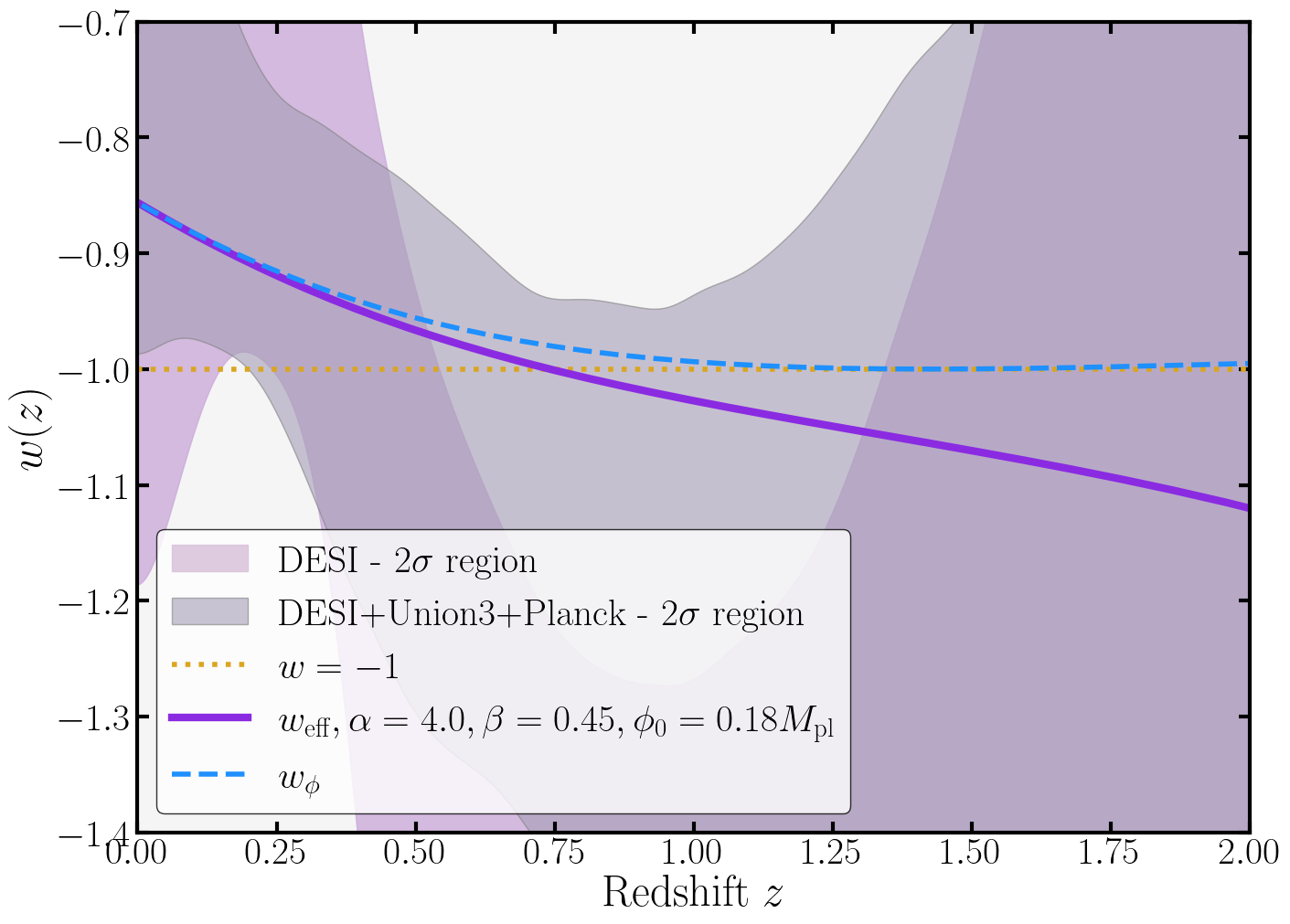}

 \includegraphics[width=0.5\linewidth]{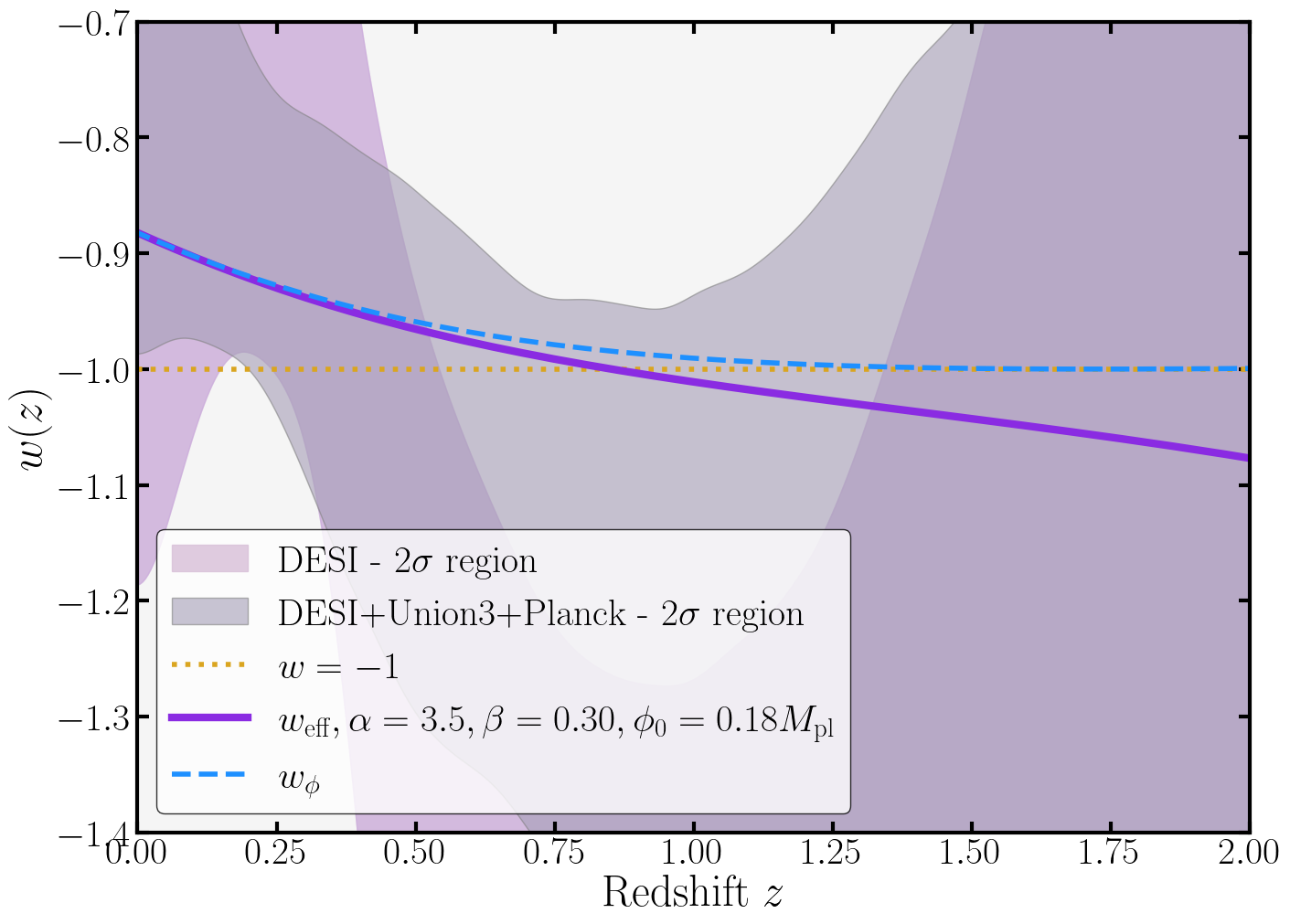}~\includegraphics[width=0.5\linewidth]{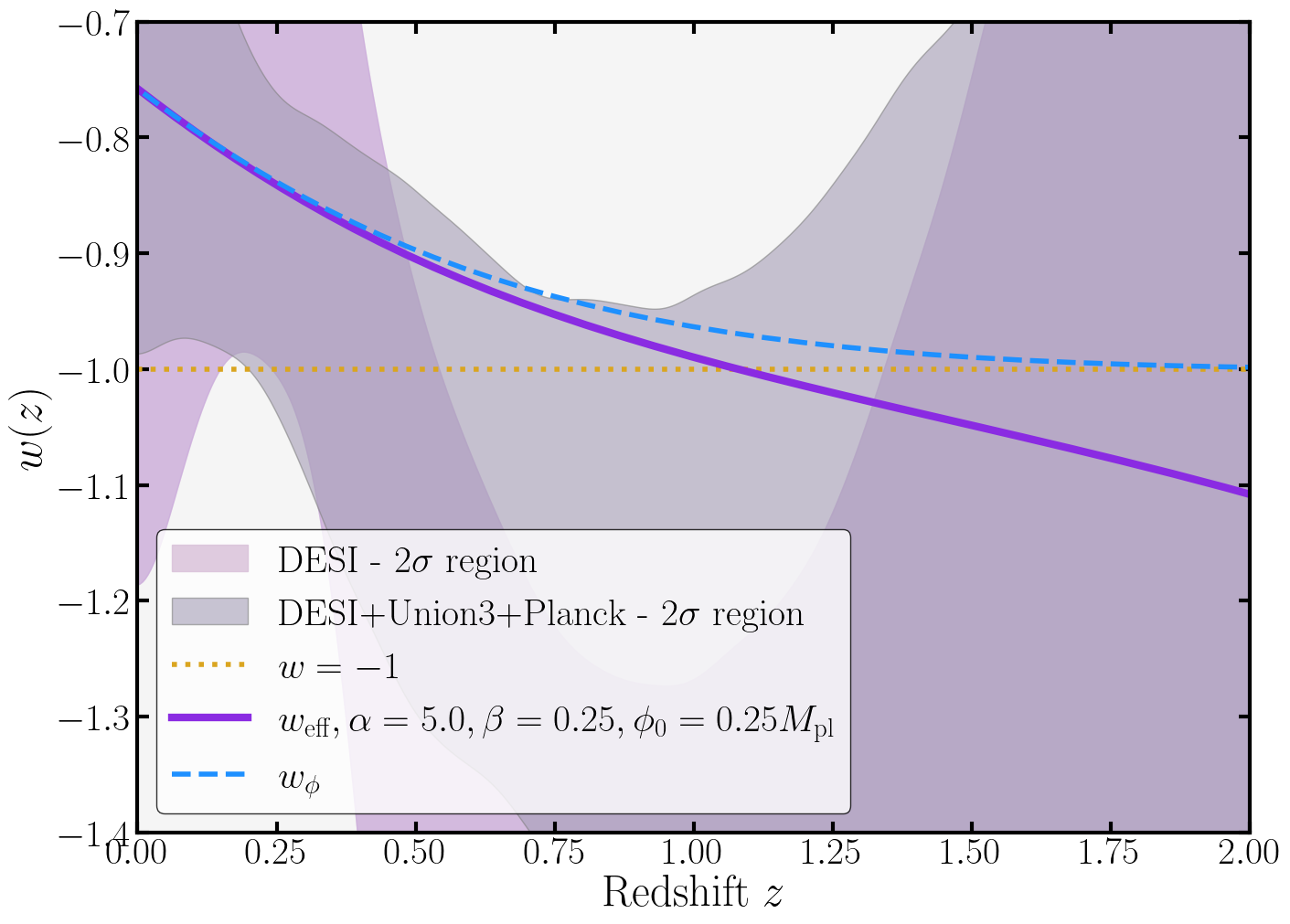}
 \caption{Fitting of $w(z)$ for exponential self-interaction potential within the $2$-$\sigma$ contours of DESI (light purple) and DESI+Union3+Planck (light gray). The initial value of phi, $\phi_{\rm in}$ is considered as $0.08 M_{\rm pl}$(top-left), $0.05 M_{\rm pl}$ (top-right), $0.05 M_{\rm pl}$ (bottom-left) and $0.025 M_{\rm pl}$ (bottom-right).}
 \label{fig:fitted_recons_exp}
\end{figure}

In the top panels of Figure \ref{fig:fitted_recons_pol}, we show predictions of the model for $\alpha=0.14, \beta=0.45$ (top-left) and $\alpha=0.14, \beta=0.40$ (top-right) using a polynomial self-interaction potential of Eq.~\eqref{eq_vphi_1} with $M=2.14\times10^{-3}~ \text{eV}$, $\phi_{\rm in} = 0.15 M_{\rm pl}$ and $\phi_0= 0.28 M_{\rm pl}$ . It is not surprising that the value of $M$ is close to the dark energy scale. We find that the prediction is not consistent with only DESI results depicted by the purple-shaded regions in the top panel. However, in the same panels, the gray-shaded regions show allowed space for the data combinations of DESI + Union3 + Planck, and we see that it fits very well within the $2-\sigma$ bounds, with $w_{\rm eff}$ crosses the phantom divide within $z \sim 0.75-1.00$.

The figures in the lower left and lower right illustrate scenarios for $\alpha =0.14$ and $\alpha =0.12$, keeping $\beta$ and $\phi_0$ fixed at $0.28$ and $0.29 M_{\rm pl}$, respectively. Both show perfect alignment within the allowed region of the combined data sets and are not consistent with DESI-only bounds, similar to the top panels. A higher value of $\alpha$ (for a fixed $\beta$) implies a higher steepness in the self-interaction potential, leading to a higher kinetic energy as the field rolls down rapidly. This behavior is reflected in $w_\phi$, where higher $\alpha$ consistently results in higher $w_\phi$ throughout all epochs presented in the plots. Consequently, $w_{\mathrm{eff}}$ also reaches a higher value in the present epoch for $\alpha=0.14$ compared to $\alpha=0.12$.
If we compare the top right and left panels of the figure where $\alpha$ remains fixed at $0.14$ and $\beta$ varies from $0.45$ to $0.40$, we find that $w_{\rm eff}$ drops sharply at higher redshifts for higher $\beta$. In this case, the dependence of $w_{\mathrm{eff}}$ on the quantity $x$, which is directly influenced by $\beta$, through Eq.~\eqref{eq:x} becomes crucial. An increase in $\beta$ leads to a higher value of $x$ at earlier epochs, with $x$ asymptotically approaching zero in the present epoch. The increase in $x$ decreases $w_{\mathrm{eff}}$. This explains the observed trend.
The most important aspect of the plots in Figure \ref{fig:fitted_recons_pol} is that we have explicit examples of a coupled dark matter quintessence model where $w_{\rm eff} < -1 $ for higher redshift with crossing $w_{\rm eff} = -1 $ at a recent time and reaching $w_{\rm eff} > -1$ at the present epoch. This is exactly the behavior that the DESI data seeks. We achieve this without any field theoretic pathologies in the quintessence dark energy sector whose equation of state always remains larger than $-1$.

We have also explored the dynamics of $w_\phi$ and $w_{\rm eff}$ for an exponential self-interaction potential of the form of Eq.~\eqref{eq_vphi_2} using $M=2.35\times10^{-3} \text{eV}$ and $\phi_0=0.18 M_{\rm pl}$ for the top and bottom left panel of the figure whereas the bottom right panel considers $\phi_0=0.25 M_{\rm pl}$. The results are illustrated in Fig.~ \ref{fig:fitted_recons_exp}. The behavior of $w_\phi$ and $w_{\rm eff}$ exhibits similar sensitivities to variations in parameters $\alpha$ and $\beta$ as in Fig.~ \ref{fig:fitted_recons_pol}. However, the order of magnitude of $\alpha$ is considerably larger for the exponential potential compared to the polynomial potential, reflecting their distinct functional forms. The steeper nature of the exponential potential enhances the kinetic energy of the scalar field, causing $w_{\rm eff}$ to exceed $-0.8$ at the current epoch while it crosses the phantom divide around $z \sim 1.00$. This has been shown in the bottom right panel of Fig.~\ref{fig:fitted_recons_exp}. 

In both Figure \ref{fig:fitted_recons_pol} and \ref{fig:fitted_recons_exp}, our result does not properly fit the DESI-only $2\sigma$ contour for any of the benchmark points considered. We don't isolate any parameter combination consistent with the DESI-only bounds. However, it's worth noting that fitting with DESI alone does not provide new insights, since the limited number of data points introduces large uncertainties in parameter inference, as reflected in the broad error bars of $w(z)$. In the era of multi-probe cosmology, priority should be given to the consistency with combined datasets -- which our model shows good agreement.

We would like to conclude the section with some comments about the nature of the potential that is suitable to fit the DESI results in this coupled scenario. Both the examples we have worked with correspond to the thawing quintessence model, where the field was stuck at its potential at some early epoch and has only started to roll recently. The equation of state behavior reported by DESI will be difficult to reproduce in a freeze-in type of dark energy model where the potential is such that the field frees into a smaller velocity as the Universe expands. In this case, the scalar field must roll down from an early redshift and come to a stop in the present day, which results in $w_\phi > -1$ at high redshift and $w_\phi \simeq -1$ at the present day. Moreover, since $\phi$ must monotonically increase with time within the observational range of DESI, it becomes impossible to implement the freezing scenario in this context. The DESI reconstruction of the equation of state excludes $w = -1$ from the $0 < z < 0.25$ range at a $2-\sigma$ confidence level for a combined Planck, DESI, and Union3 dataset \cite{Reconstruction}. Due to the expected smooth behavior of scalar field dynamics, the coupling function $f(\phi)), x, w_\phi$, and $w_{\rm eff}$ all exhibit smoothness as well. At present, when $x = 0$ , we have $w_\phi = -1$ , which implies that $w_{\rm eff}= -1$ . To maintain the smooth behavior of $w_{\rm eff}$, it would be impossible for $w_{\rm eff}$ to exceed -1 for the redshift range of $0 < z < 0.25$ for the combined dataset. Therefore, a freezing scenario with interaction is unlikely to adequately fit the DESI results in its simplest form. 

\section{Conclusions and Outlook}\label{sec:conclusions}

In this study, we explore a scenario of a quintessence dark energy paradigm coupled with dark matter inspired by the framework proposed by \cite{Das:2005yj}. A key feature of our approach is a scalar field that moves away from its minimum with significant kinetic energy in the present epoch, leading to a dynamical evolution of dark energy that effectively includes phantom crossing\footnote{The effects of kinetic energy in explaining DESI results have been discussed in \cite{Berghaus:2024kra} }. This approach naturally accommodates the crucial observational constraints, especially from the Dark Energy Spectroscopic Instrument (DESI) survey. The distinctive capability of our model to cross the phantom divide, with the effective equation of state parameter $w_{\mathrm{eff}}$ transitioning from values below $-1$ in the distant past to values exceeding $-1$ in the present epoch, offers a compelling explanation for the observed signs of dark energy evolution. A crucial difference in comparison to many other attempts to explain the DESI results is that the dark energy sector is devoid of any pathologies, with the equation of state parameter $w_{\phi} > -1$ always. 

Our investigation is further enhanced by numerically solving the modified Klein-Gordon equation for two distinct self-interaction potentials, namely exponential and polynomial, allowing us to explore the scalar field dynamics and compute the effective equation of state, $w_{\mathrm{eff}}$. The results, particularly when compared across various combinations of the parameters $\alpha$ and $\beta$, show strong agreement with data from the combined DESI, Planck, and Union3 datasets. We found that the set-up has enough flexibility for the $w_{\rm eff}$ to cross the phantom-divide for a range of $z$ and also its present value reaching to be as large as $w_{\rm eff} \sim -0.8$.

This work provides a novel step toward a fundamental understanding of the complex interplay between dark energy and dark matter. Using current DESI data, we have shown that our framework offers a viable description, while a comprehensive Markov Chain Monte Carlo (MCMC) study remains to more precisely map the parameter space and to further improve the statistical robustness of the results. The continued release of DESI observations will be crucial to this effort, helping to close present gaps and to shed further insights on the dynamics of dark energy. In our setup, the dark energy field couples only to dark matter, and we have assumed that quantum corrections from the dark matter sector—being heavier than the dark energy mass—do not modify the latter’s mass. Embedding these requirements in a concrete particle-physics model is nontrivial and will be pursued in future work.

Moreover, the proposed additional interactions in the dark matter-dark energy fluid naturally give rise to
modified perturbation equations, which may lead to observational consequences in the CMB and LSS data. We have discussed the perturbation equations in the Appendix and have argued that the effect on the perturbations will be small and should be within the observational bounds. But a full cosmological analysis, including perturbations, is crucial to validate our model, and that is actively being pursued now.

Overall, this work contributes to the growing body of research aimed at unraveling the mysteries of dark energy and its role in the accelerated expansion of the universe. By offering a model that naturally accommodates evolving dark energy and phantom crossing, we pave the way for further investigations into alternative cosmological models that may better align with future observational data.\\

\noindent Note added: Soon after we posted our work on the arXiv, the DESI DR2 dataset was released \cite{DESI:2025zgx}, which reconfirms all the findings of their previous analyses on which our study is based. We do not anticipate any substantial changes to our conclusions in light of the new data release. 

\section{Appendix: Perturbations for dark matter-dark energy fluid }\label{sec:perturb}
The presence of an additional interaction in the dark matter-dark energy fluid naturally gives rise to modified perturbation equations, whose equation of motion is presented as follows,

\begin{equation} 
\begin{aligned}
\dot{\delta}_{\textrm{\tiny DM}}&=-\left( \frac{\theta_{\textrm{\tiny DM}}}{a} + 
\frac{\dot{\tilde{h}}}{2} \right) + \frac{\tilde{\beta}}{M_\textrm{\tiny Pl}} \delta \dot{\phi} \ , \\
\dot{\theta}_{\textrm{\tiny DM}} &= -H\theta_{\textrm{\tiny DM}}+\frac{\tilde{\beta}}{M_\textrm{\tiny Pl}}\left( \frac{k^2}{a}\delta\phi -\dot{\phi}\theta_{\textrm{\tiny DM}} \right) \ , \label{eq:coldvelocity}
\end{aligned}
\end{equation} 
Where, $\tilde{\beta}=\frac{\beta}{\sqrt{8\pi}}$. Also, the perturbed Klein-Gordon equation becomes,
\begin{equation} 
\ddot{\delta \phi} + 3H\dot{\delta \phi}+\left( \frac{k^2}{a^2} +V_{,\phi\phi} \right)\delta\phi + \frac{1}{2}\dot{\tilde{h}}\dot{\phi}=-\frac{\tilde{\beta}}{M_\textrm{\tiny Pl}} \rho_{\textrm{\tiny DM}}\delta_{\textrm{\tiny DM}} \ ,  \label{eq:scalarperturbations}
\end{equation}  
where $\tilde{h} \equiv \tilde{h}^i_{\ i}$ is the trace of metric perturbation~$\tilde{h}_{ij}$ in synchronous gauge.

Similarly, the perturbed Einstein equation determining the evolution of metric perturbation gets modified too in the following manner,
\begin{eqnarray}
\label{eq:metricperturb}
 k^2 \eta -\frac{1}{2} a^2 H \dot{\tilde{h}} &=& -\frac{a^2}{2 M^2_\textrm{\tiny Pl}} \left(\sum_{i=\gamma,\nu,\textrm{\tiny B},\textrm{\tiny DM}}
\rho_i \delta_i +  \delta \rho_\phi \right), \\
 k^2 \dot{\eta} &=& \frac{a}{2 M^2_\textrm{\tiny Pl}} \left( \sum_{i=\gamma,\nu,\textrm{\tiny B},\textrm{\tiny DM}} (\rho_i+ P_i) \theta_i + 
a k^2 \dot{\phi} \delta \phi \right) \ , \\
\ddot{\tilde{h}} + 3 H  \dot{\tilde{h}} - 2 \frac{k^2}{a^2} \eta & =&  - \frac{3}{M_\textrm{\tiny Pl}^2} \left(\sum_{i=\gamma,\nu,\textrm{\tiny B},\textrm{\tiny DM}}
 \delta P_i  + \dot{\phi}\delta\dot{\phi}-V_{,\phi}\delta\phi  \right), \label{eq:metricperturb2}
\end{eqnarray}
where $\delta \rho_\phi = \dot{\phi}\delta\dot{\phi}+V_{,\phi}\delta\phi$.

Since the field behaves similarly to a thawing quintessence model, but with dynamics governed by an effective potential---where the field begins to roll only at very late times ($z \sim 4$)---the kinetic energy of the field remains subdominant to its potential energy. This is evident in the evolution of $w_{\phi}$, shown later in Figures~\ref{fig:fitted_recons_pol} and \ref{fig:fitted_recons_exp} in Section~\ref{Sec:result}, where the equation of state does not vary significantly, changing from $-1$ to approximately $-0.8$ within the redshift range $0$--$4$. It can be shown that the kinetic energy remains smaller than the potential energy as long as $w_{\phi}$ stays negative. Therefore, in the perturbation equation, we can safely consider $\dot{\phi}$ to be small and assume $\ddot{\delta \phi} = \dot{\delta \phi} = 0$. The perturbation equation for the growth of dark matter then becomes,

\begin{equation}
\label{eq:delta_ptb_vel}
\ddot{\delta}_\textrm{\tiny DM} + \left(2 H + \frac{\tilde{\beta}\dot{\phi}}{M_\textrm{\tiny Pl}}\right) \dot{\delta}_\textrm{\tiny DM} + \frac{1}{2} \left( \ddot{\tilde{h}} + \left(2 H + \frac{\tilde{\beta}\dot{\phi}}{M_\textrm{\tiny Pl}}\right)\dot {\tilde{h}} \right) \approx \frac{\tilde{\beta}^2}{M_\textrm{\tiny Pl}^2} \frac{\rho_\textrm{\tiny DM} \delta_\textrm{\tiny DM}}{1+a^2 V_{,\phi \phi}/k^2} +  \frac{\frac{1}{2}\frac{\tilde{\beta}\dot{\phi}}{M_\textrm{\tiny Pl}}\dot{\tilde{h}}}{1+a^2 V_{,\phi \phi}/k^2}\ ,
\end{equation}

We kept the terms proportional to $\dot{\phi}$, to quantify the effect on the growth of perturbation when the scalar field has a small velocity. Here, we ignore other components in the perturbation equation for simplicity. 

Now initially, the field is stuck at the initial point and only starts to roll from redshift $z\sim 4$. So before that, the perturbation growth follows the following equation,

\begin{equation}
\label{eq:delta_ptb_nor}
\ddot{\delta}_\textrm{\tiny DM} + 2 H \dot{\delta}_\textrm{\tiny DM} + \frac{1}{2} \left( \ddot{\tilde{h}} + 2 H \dot {\tilde{h}} \right) \approx \frac{\tilde{\beta}^2}{M_\textrm{\tiny Pl}^2} \frac{\rho_\textrm{\tiny DM} \delta_\textrm{\tiny DM}}{1+a^2 V_{,\phi \phi}/k^2}\ ,
\end{equation}

The right-hand side term arises due to the extra fifth force in the system. The denominator of this term quantifies the screening of this force, which shows the comoving range of this force $\lambda_\textrm{\tiny F} (a) \equiv a^{-1} V_{,\phi \phi}^{-1/2}$. So only the modes $k > \lambda_\textrm{\tiny F}^{-1}$ will be affected by this extra force, in turn enhancing the growth at those scales. The perturbation growth of the modes till $z\sim 4$ is similar to the growth shown and argued in \cite{Das:2005yj, Boriero:2015loa}.

Now, after redshift $z=4$, the field starts rolling and starts gaining kinetic energy. So this scenario can be thought of as a correction to the growth dynamics discussed in \cite{Das:2005yj, Boriero:2015loa}. Comparing Equation \ref{eq:delta_ptb_vel} with \ref{eq:delta_ptb_nor}, we see some extra contributions are coming from the velocity of the field. As argued before, the field velocity remains comparatively small throughout the field evolution, suggesting a small correction to the dynamics in the growth of the perturbation. 

The L.H.S of Equation \ref{eq:delta_ptb_vel}, shows an extra damping term proportional to $\tilde{\beta}\dot{\phi}$, which shows an overall supression at scales of dark matter perturbation, however, $\frac{\tilde{\beta}\dot{\phi}}{M_\textrm{\tiny Pl}}\ll 2H$, due to both $\beta<0.5$ and velocity of the field being small. So we expect the extra suppression due to field velocity to be very tiny. Following the same argument, we can also infer that the expression $\left( \ddot{\tilde{h}} + \left(2 H + \frac{\tilde{\beta}\dot{\phi}}{M_\textrm{\tiny Pl}}\right)\dot {\tilde{h}} \right)$ predominantly should depend on $\propto -\rho_\textrm{\tiny DM} \delta_\textrm{\tiny DM}$, which acts as a source term for the growth of the perturbation. 

In the R.H.S of Equation \ref{eq:delta_ptb_vel}, there is an extra velocity-dependent source term in addition to the standard fifth force source term. Now this term also acts source of growth of perturbation but within the co-moving range of the fifth force $\lambda_\textrm{\tiny F}$. So it enhances the growth for scales $k > \lambda_\textrm{\tiny F}^{-1}$. We can gain some intuition regarding the physical in this type of coupled dark sector model, through analytically deriving the perturbation equation of growth of perturbation equation under some assumptions. However, to truly get an idea of how different modes get affected, one should implement the model in a Boltzmann code \texttt{CLASS} to generate the matter power spectrum, which is beyond the scope of this paper, but we keep it for our future study.

\acknowledgments
We thank Shadab Alam, Eric Linder, and Anton Chudaykin for the helpful discussion and useful comments regarding the manuscript. We also thank Xin Wang for pointing out an important typo in the texts.

\bibliographystyle{JHEP}
\bibliography{main.bib}

\end{document}